\documentclass[aps,prl,english,amsmath,amsfonts,amssymb,superscriptaddress,twocolumn,showpacs,citeautoscript,reprint,longbibliography]{revtex4-1}
\usepackage[T1]{fontenc}
\usepackage[latin9]{inputenc}
\usepackage{amsmath}
\usepackage{amssymb}
\usepackage{wasysym}
\usepackage{adjustbox}
\usepackage{esint}
\usepackage{graphicx}
\usepackage{times}
\usepackage{nicefrac}
\usepackage{appendix}
\usepackage{multirow}
\usepackage{textcase}
\usepackage{subfigure}
\usepackage{empheq}
\usepackage{color}
\usepackage{leftidx}
\usepackage{hhline}
\usepackage{makecell}
\usepackage{stackrel}
\usepackage{lineno}
\makeatletter
\@ifundefined{textcolor}{}
{%
 \definecolor{BLACK}{gray}{0}
 \definecolor{WHITE}{gray}{1}
 \definecolor{RED}{rgb}{1,0,0}
 \definecolor{GREEN}{rgb}{0,1,0}
 \definecolor{BLUE}{rgb}{0,0,1}
 \definecolor{CYAN}{cmyk}{1,0,0,0}
 \definecolor{MAGENTA}{cmyk}{0,1,0,0}
 \definecolor{YELLOW}{cmyk}{0,0,1,0}
}

\renewcommand{\vec}[1]{\mathbf{#1}}
\renewcommand{\Re}{\operatorname{Re}}
\renewcommand{\Im}{\operatorname{Im}}
\newcommand{\tr}{\operatorname{Tr}}

\newcommand{\asym}{\operatorname{asym}}

\renewcommand{\b}{\beta}

\newcommand{\Phib}{\Phi_\mathrm{Born}}

\newcommand{\add}[1]{\if\a\b{{\color{red} #1}}\else{#1}\fi}

\newcommand{\bracket}[1]{\langle #1 \rangle}
\newcommand{\ket}[1]{| #1 \rangle}
\newcommand{\bra}[1]{\langle #1 |}
\newcommand{\im}{i}

\renewcommand{\eqref}[1]{(\ref{eq:#1})}

\newcommand{\figref}[1]{Fig.~\ref{fig:#1}}

\newcommand{\TT}{\mathbb{T}}
\newcommand{\VV}{\mathbb{V}}
\newcommand{\GG}{\mathbb{G}}

\newcommand{\II}{\mathbb{I}}
\newcommand{\OO}{\mathbb{O}}
\newcommand{\DD}{\mathbb{D}}
\newcommand{\Gvac}{\mathbb{G}^{\mathrm{vac}}}

\makeatother
\usepackage{babel}

\begin{document}
\title{Fundamental limits to radiative heat transfer: the limited role
  of nanostructuring in the near-field}

\author{Prashanth S. Venkataram}
\author{Sean Molesky}
\author{Weiliang Jin}
\author{Alejandro W. Rodriguez}
\affiliation{Department of Electrical Engineering, Princeton
  University, Princeton, New Jersey 08544, USA}

\date{\today}

\begin{abstract}
  In a complementary article~\cite{MoleskyARXIV2019A}, we exploited
  algebraic properties of Maxwell's equations and fundamental
  principles such as electromagnetic reciprocity and passivity, to
  derive fundamental limits to radiative heat transfer applicable in
  near- through far-field regimes. The limits depend on the choice of
  material susceptibilities and bounding surfaces enclosing
  arbitrarily shaped objects.  In this article, we apply these bounds
  to two different geometric configurations of interest, namely
  dipolar particles or extended structures of infinite area in the
  near field of one another, and compare these predictions to prior
  limits. We find that while near-field radiative heat transfer
  between dipolar particles can saturate purely geometric ``Landauer''
  limits, bounds on extended structures cannot, instead growing much
  more slowly with respect to a material response figure of merit, an
  ``inverse resistivity'' for metals, due to the deleterious effects
  of multiple scattering; nanostructuring is unable to overcome these
  limits, which can be practically reached by planar media at the
  surface polariton condition.
\end{abstract}

\maketitle 

Radiative heat transfer (RHT) between two bodies may be written as a
frequency integral of the form
\begin{equation} \label{eq:integratedRHT}
  P = \int_{0}^{\infty} [\Pi(\omega, T_{\mathrm{B}}) -
  \Pi(\omega, T_{\mathrm{A}})]\Phi(\omega)~\mathrm{d}\omega
\end{equation}
where $\Pi(\omega, T)$ is the Planck function (and it has been
assumed, without loss of generality, that $T_{\mathrm{B}} >
T_{\mathrm{A}}$ so $P > 0$), and $\Phi(\omega)$ a dimensionless
spectrum of energy transfer. RHT between two objects sufficiently
separated in space follows the Planck blackbody law, but in the
near-field where separations are smaller than the characteristic
thermal wavelength of radiation, contributions to RHT from evanescent
modes will dominate, allowing $\Phi(\omega)$ to exceed the far-field
blackbody limits by orders of magnitude. Moreover, because the Planck
function decays exponentially with frequency, judicious choice of
materials and nanostructured geometries can shift resonances in $\Phi$
to lower (especially infrared) frequencies, allowing observation of
even larger integrated RHT powers~\cite{VolokitinPRB2001,
  DominguesPRL2005, VolokitinRMP2007, SongNATURENANO2015}. However,
after accounting for the effects of such frequency shifts, the degree
to which the spectrum $\Phi$ at a given frequency can be enhanced
remains an open question. The inability of trial-and-error
explorations and optimization procedures~\cite{JinOE17,
  FernandezHurtadoPRL2017} to saturate prior bounds on $\Phi$ based on
modal analyses~\cite{PendryJPCM1999, BimontePRA2009, BiehsPRL2010,
  BenAbdallahPRB2010} or energy conservation~\cite{MillerPRL2015}
suggests that these prior bounds may be too loose.

In a complementary article~\cite{MoleskyARXIV2019A}, we derived new
bounds that simultaneously account for material and geometric
constraints as well as multiple scattering effects. These bounds,
valid from the near- through far-field regimes, incorporate the
dependence of the optimal modal response of each object on the other
while simultaneously being constrained by passivity considerations in
isolation. They depend on a general material response factor
(``inverse resistivity'' for metals)~\cite{MillerPRL2015},
\begin{equation}
  \zeta = \frac{|\chi|^{2}}{\Im(\chi)},
\end{equation}
without making explicit reference to specific frequencies or
dispersion models, and are domain monotonic, increasing with object
volumes independently of their shapes. Consequently, our bounds are
applicable at all length scales, from quasistatic to ray optics
regimes, do not suffer from unphysical divergences with respect to
vanishing material dissipation or object sizes~\cite{MillerPRL2015},
and can be interpreted independently of specific object shapes.

In this article, we apply the aforementioned bounds on $\Phi$ to two
geometric configurations of practical interest, comparing predictions
to prior bounds based on energy conservation~\cite{MillerPRL2015},
applicable only in the quasistatic regime, or Landauer-like modal
summations~\cite{PendryJPCM1999, BimontePRA2009, BiehsPRL2010,
  BenAbdallahPRB2010}, applicable only in the ray optics
regime. Specifically, we consider limits on RHT between dipolar
particles as well as extended structures of infinite area and
arbitrary shapes restricted to the \emph{near field}. We find that our
exact bound for dipolar particles is able to reach Landauer limits
when $\zeta$ exceeds a certain threshold; in contrast, bounds that
neglect losses due to multiple scattering grossly overestimate
possible material enhancements, diverging with increasing $\zeta$. For
extended structures, we find that the bound grows only weakly
(logarithmically) with respect to $\zeta$, making the neglect of
multiple scattering even more apparent. Fundamentally, previous
limits~\cite{MillerPRL2015} were based on a Born approximation which,
in analogy with Kirchhoff's law~\cite{VolokitinPRB2001,
  VolokitinRMP2007}, assumed that thermal fields produced within a
given body in isolation can be perfectly absorbed by others in
proximity. This explains the aforementioned performance gap: the
combination of resonant absorption and multiple scattering hampers
rather than helps NFRHT, and the previous bounds cannot capture this
trade-off. Finally, we discuss practical implications and design
guidelines for structures enhancing NFRHT.

\begin{table}[t] 
\label{tab:summary}
\centering
\begin{adjustbox}{center, width=\columnwidth-10pt}
  \begin{tabular}{|c|c|c|c|}
    \hline\rule{0pt}{4.ex}Bound & Formula & %Material & Multiple
    \hspace{-0.05in} $\begin{array}{cc} \text{Material} &
      \\ \text{factor} \end{array}$ \hspace{-0.15in} &
    \hspace{-0.05in} $\begin{array}{cc} \text{Multiple} &
      \\ \text{scattering} \end{array}$ \hspace{-0.15in}
    \\[1.7ex] \hhline{|=|=|=|=|} \rule{0pt}{2.6em} $\Phi_\mathrm{opt}$
    & \hspace{-0.1in}$\begin{array}{lcl} && \sum_{i} \frac{1}{2\pi}
      \Theta(\zeta_{\mathrm{A}} \zeta_{\mathrm{B}} g_{i}^{2} - 1)
      \\ &+&\sum_{i} \frac{2}{\pi} \frac{\zeta_{\mathrm{A}}
        \zeta_{\mathrm{B}} g_{i}^{2}}{(1 + \zeta_{\mathrm{A}}
        \zeta_{\mathrm{B}} g_{i}^{2})^{2}} \Theta(1 -
      \zeta_{\mathrm{A}} \zeta_{\mathrm{B}} g_{i}^{2}) \end{array}$ &
    Yes & Yes \\[1.7em] \hline \rule{0pt}{1.5em}
    $\Phi_\mathrm{Born}$ & $\sum_{i} \frac{2}{\pi} \zeta_{\mathrm{A}}
    \zeta_{\mathrm{B}} g_{i}^{2}$ & Yes & No \\[1.5ex] \hline
    \rule{0pt}{1.5em} $\Phi_\mathrm{L}$ & $\sum_{i} \frac{1}{2\pi}$ &
    No & No \\[1.5ex] \hline \rule{0pt}{1.5em}$\Phi_\mathrm{sc}$ &
    $\sum_{i} \frac{2}{\pi} \frac{\zeta_{\mathrm{A}}
      \zeta_{\mathrm{B}} g_{i}^{2}}{(1 + \zeta_{\mathrm{A}}
      \zeta_{\mathrm{B}} g_{i}^{2})^{2}}$ & Yes & Yes \\[.8em] \hline
  \end{tabular}
\end{adjustbox}
\caption{\textbf{Summary of various bounds on NFRHT.}
  $\Phi_{\mathrm{opt}}$ captures multiple scattering and geometric
  constraints via the singular values $\{g_{i}\}$ of the vacuum
  Green's function $\Gvac_{\mathrm{BA}}$, and material constraints via
  the response factors $\zeta_{p} =
  \frac{|\chi_{p}|^2}{\Im(\chi_{p})}$ for $p=\{\mathrm{A},
  \mathrm{B}\}$. $\Theta$ is the Heaviside step function. As described
  in the main text, restricted versions of $\Phi_{\mathrm{opt}}$ each
  capture different facets of this bound.}
\end{table}

\emph{General bounds.---}We now briefly recapitulate the bounds on RHT
between bodies A and B derived in~\cite{MoleskyARXIV2019A} and
describe their salient features; readers may
follow~\cite{MoleskyARXIV2019A} for more technical details. These
bounds are derived for bodies $p \in \{\mathrm{A}, \mathrm{B}\}$ with
arbitrary homogeneous local isotropic susceptibilities $\chi_{p}$ and
arbitrary shape and size. They depend on material constraints,
particularly passivity (nonnegativity of far-field scattering by each
object in isolation and in the presence of the other), encoded in the
response factors $\zeta_{p} = |\chi_{p}|^{2} / \Im(\chi_{p})$, and on
geometric constraints encoded in the off-diagonal vacuum Maxwell
Green's function $\Gvac_{\mathrm{BA}}$, which solves $[(c/\omega)^{2}
  \nabla \times (\nabla \times) - \II]\Gvac = \II$. In particular, the
bounds rest on the singular values $\{g_{i}\}$ obtained from a
singular-value decomposition,
\begin{equation}
  \Gvac_{\mathrm{BA}} = \sum_{i} g_{i} \ket{\vec{b}_{i}} \bra{\vec{a}_{i}},
\end{equation}
where $\ket{\vec{a}_{i}}$ and $\ket{\vec{b}_{i}}$ are the
corresponding right and left singular vectors, respectively. A key
property of this expansion is that the singular values of
$\Gvac_{\mathrm{BA}}$ are domain-monotonic, increasing with increasing
domain volume.

We list the relevant bounds in Table 1. The main results of this paper
rely on the upper bound $\Phi_{\mathrm{opt}}$, which we refer to as an
``exact bound'' in that it is valid from the near- through far-field
regimes, though below we focus only on near-field
effects. $\Phi_\mathrm{opt}$ is domain monotonic in that it always
increases with increasing object volumes, and this comes from the
domain monotonicity of $g_{i}$. Therefore, one can choose to evaluate
the bound in a domain of high symmetry enclosing the objects of
interest, representing a fundamental geometric constraint in analogy
and in combination with material constraints imposed by a specific
choice of $\zeta_{p}$.

The expression for $\Phi_{\mathrm{opt}}$ makes clear that optimal heat
transfer is achievable if the modes of the response of each body
coincide with the modes of the \emph{vacuum} Green's function
$\Gvac_{\mathrm{BA}}$. Additionally, for each channel $i$, each term
may be physically interpreted as follows. The first term
$\frac{1}{2\pi}$ corresponds to the Landauer limit for that channel,
which is the maximum possible contribution to $\Phi$ for a given
channel~\cite{Datta1995, KlocknerPRB2016, PendryJPCM1999,
  BimontePRA2009, BiehsPRL2010, BenAbdallahPRB2010}; a given channel
$i$ attains this only if $\zeta_{\mathrm{A}} \zeta_{\mathrm{B}}
g_{i}^{2} \geq 1$, meaning that while channels that efficiently couple
electromagnetic fields propagating in vacuum between the two bodies
can lead to saturation, channels that do not require instead larger
material response factors $\zeta_{p}$. In contrast, the total Landauer
bound $\Phi_{\mathrm{L}}$ assumes saturation of every channel $i$ (the
first term) regardless of material response or geometric
configuration. The second term $\frac{2}{\pi} \frac{\zeta_{\mathrm{A}}
  \zeta_{\mathrm{B}} g_{i}^{2}}{(1 + \zeta_{\mathrm{A}}
  \zeta_{\mathrm{B}} g_{i}^{2})^{2}}$, which never exceeds the
per-channel Landauer limit of $\frac{1}{2\pi}$, corresponds to each
body attaining its maximum absorptive response in isolation for the
respective incident fields $\ket{\vec{a}_{i}}$ and $\ket{\vec{b}_{i}}$
for channel $i$ in order to satisfy passivity constraints; the
numerator corresponds to the contribution from absorption of each body
in isolation, while the denominator captures multiple scattering
effects among bodies. In contrast, the ``scalar approximation''
$\Phi_{\mathrm{sc}}$ assumes that each body exhibits maximal isolated
absorption (i.e. uniform or scalar response) corresponding to the
second term for every channel $i$; while $\Phi_{\mathrm{sc}}$ includes
both material response constraints in the numerator and multiple
scattering effects in the denominator, the ``Born bound''
$\Phi_{\mathrm{Born}}$ further dispenses with the denominator
(i.e. multiple scattering effects) entirely for every channel
$i$~\cite{MillerPRL2015}. In \cite{MoleskyARXIV2019A}, we proved that
these bounds satisfy the inequalities
\begin{equation} \label{eq:Phirelations}
  \Phi_{\mathrm{sc}} \leq \Phi_{\mathrm{opt}} \leq
  \Phi_{\mathrm{Born}}, \Phi_{\mathrm{L}}
\end{equation}
regardless of the particular bounding domain, and thus we may compare
them for specific topologies of interest.

\begin{figure}[t!]
\centering
\includegraphics[width=0.9\columnwidth]{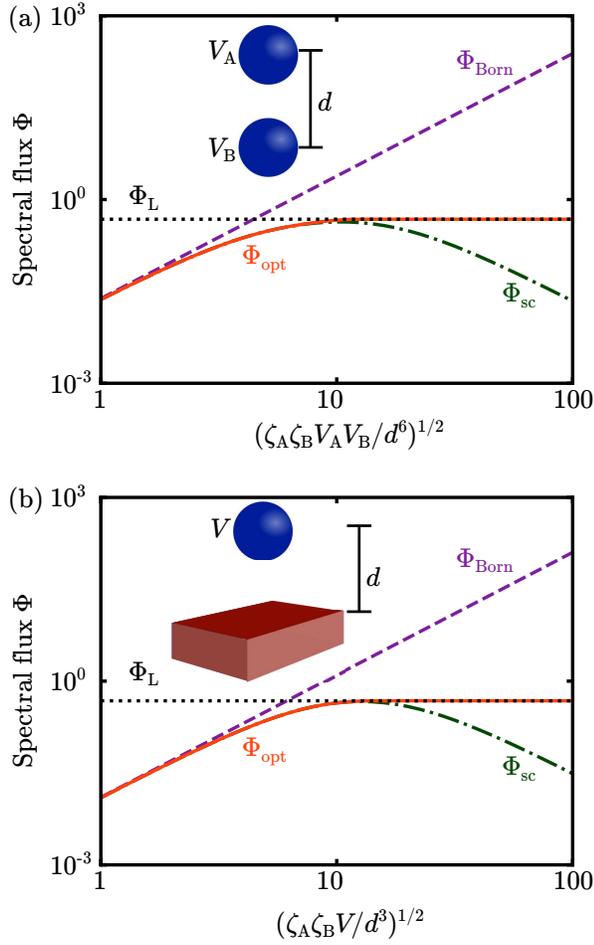}
\caption{Comparison of $\Phi_{\mathrm{opt}}$ (solid orange) to
  $\Phi_{\mathrm{L}}$ (dotted black), $\Phi_{\mathrm{Born}}$ (dashed
  purple), and $\Phi_{\mathrm{sc}}$ (dot-dashed green) for a dipolar
  body separated by distance $d$ from (a) another dipolar body, in
  which case both dipolar volumes $V_{\mathrm{A}}$ and
  $V_{\mathrm{B}}$ are relevant, or (b) an extended structure, in
  which case only the single dipolar volume $V$ is relevant.}
\label{fig:dipolebounds}
\end{figure}

\emph{Dipolar bodies.---} We first consider NFRHT between either two
dipolar particles [\figref{dipolebounds}(a)] or a dipolar particle and
an extended bulk medium of infinite area and thickness
[\figref{dipolebounds}(b)], enclosed within spherical or semi-infinite
bounding domains, as detailed in the appendices. The dipolar limit
implies that if $V$ is the volume of a dipolar particle and $d$ is the
separation from the other body, then $\frac{V^{1/3}}{d} \ll 1$, and no
higher-order particle multipoles should matter. This also implies
that there are only 3 degrees of freedom or singular values
(i.e. polarizations) and therefore 3 channels of interest, meaning
that in either case, we can immediately write the Landauer limit as
$\Phi_{\mathrm{L}} = \frac{3}{2\pi}$. As we show in the appendices, in
the first case, the quantities $\Phi_{\mathrm{opt}}$,
$\Phi_{\mathrm{Born}}$, and $\Phi_{\mathrm{sc}}$ depend only on the
combined quantity $\frac{\sqrt{\zeta_{\mathrm{A}} \zeta_{\mathrm{B}}
    V_{\mathrm{A}} V_{\mathrm{B}}}}{d^{3}}$ where $V_{p}$ is the
volume of each dipolar body $p \in \{\mathrm{A}, \mathrm{B}\}$, while
in the second case, they depend on $\sqrt{\frac{\zeta_{\mathrm{A}}
    \zeta_{\mathrm{B}} V}{d^{3}}}$ where $V$ is the volume of the one
dipolar body.

In both cases, the Born bound depends linearly on the product
$\zeta_{\mathrm{A}} \zeta_{\mathrm{B}}$, which explains why for
increasing material factors (assuming fixed volumes and separations)
the bound eventually crosses the Landauer limits. By contrast,
$\Phi_{\mathrm{opt}}$ will never cross or exceed either Landauer or
Born bounds, while hugging the latter from below and increasing
monotonically toward $\Phi_\mathrm{L}$ with increasing material
factors (e.g. small dissipation). We note that whether the dipolar
particle is near another or an extended structure, the smallest two
singular values of $\Gvac_{\mathrm{AB}}$ are equal to each other and
correspond to the two axes perpendicular to the line of separation,
while the largest singular value is larger than the smaller two by
different factors depending on the particular case. This dependence
therefore implies that for the Landauer bounds to be saturated, the
optimal net response of each body cannot be isotropic, even though the
underlying susceptibilities are assumed to be isotropic; the optimal
dipole should instead arise for an oblate ellipsoidal shape whose
aspect ratio is a function of $g_{\max}/g_{\min}$, while the optimal
extended structure (assuming an isotropic particle) should be textured
in order to break homogeneity. The scalar approximation in each case
hugs $\Phi_{\mathrm{opt}}$ from below up until it smoothly reaches a
peak, and then decays as a power law thereafter. The peak value of
$\Phi_{\mathrm{sc}}$ is within 10\% of the Landauer bound in each
case, suggesting that for susceptibilities and frequencies chosen to
give an appropriate value of $\zeta_{\mathrm{A}} \zeta_{\mathrm{B}}$,
the limits can practically be reached by isotropic spherical dipoles
and thick planar films; we note that the surface polariton condition
is $\Re(1/\chi) = -1/2$ for a planar film, or $\Re(1/\chi) = -1/3$ for
a dipolar sphere. However, the assumption of maximum isolated
absorption implies that for $\zeta_{\mathrm{A}} \zeta_{\mathrm{B}}$
larger than the aforementioned threshold, $\Phi_{\mathrm{sc}}$ is a
local \emph{minimum} rather than maximum and starts decreasing with
respect to $\zeta_{\mathrm{A}} \zeta_{\mathrm{B}}$ as multiple
scattering becomes deleterious for such configurations; such is the
price of approximating and restricting the response of the system to
be uniform instead of allowing the response to vary per channel.

\begin{figure}[t!]
\centering
\includegraphics[width=0.9\columnwidth]{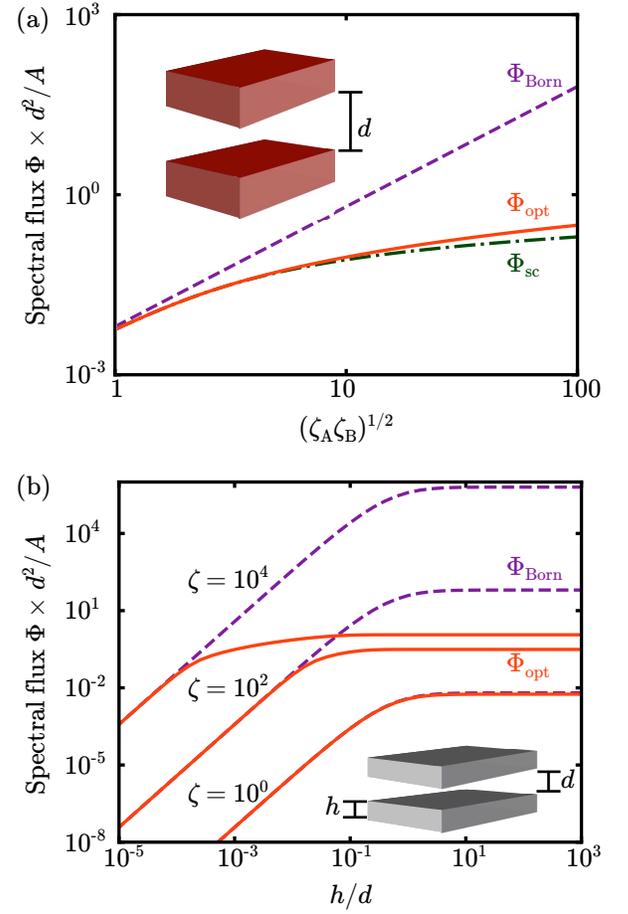}
\caption{Comparison of $\Phi_{\mathrm{opt}}$ (solid orange) to
  $\Phi_{\mathrm{Born}}$ (dashed purple) and $\Phi_{\mathrm{sc}}$
  (dot-dashed green) for two extended structures of infinite area $A$
  and (a) infinite thickness or (b) finite thickness $h$ normalized to
  their mutual separation $d$.  Both plots illustrate the behavior of
  $\Phi$ (normalized by $A/d^{2}$) with respect to material factors;
  $\Phi_{\mathrm{sc}}$ is not shown in (b) due to the near-overlap
  with $\Phi_{\mathrm{opt}}$.}
\label{fig:platebounds}
\end{figure}

\emph{Extended structures.---} We now consider NFRHT between two
extended structures of infinite area $A$ separated by a distance
$d$. In this case, there is an infinite continuum of channels that may
participate, labeled by the two-dimensional in-plane wavevector
$\vec{k}$, and the sum over channels $i$ is written $\sum_{i} \to
A\iint \frac{\mathrm{d}^{2} k}{(2\pi)^{2}}$. Furthermore, even after
normalizing to the area, the Landauer bound $\Phi_{\mathrm{L}}/A =
\iint \frac{1}{2\pi} \frac{d^{2} k}{(2\pi)^{2}}$ diverges, so we do
not consider it further, and instead only consider
$\Phi_{\mathrm{opt}}$, $\Phi_{\mathrm{Born}}$, and
$\Phi_{\mathrm{sc}}$ after multiplying by a common factor of
$\frac{d^{2}}{A}$, each of which only depend on the product of
material factors $\sqrt{\zeta_{\mathrm{A}} \zeta_{\mathrm{B}}}$ and on
no other length scales in the near-field.

As we show in the appendices, for two planar semi-infinite half-spaces
constituting the bounding regions, these bounds take on particularly
simple analytical forms, with
\begin{align}
\label{eq:Phiopt_ext}
%  \begin{split}
    \Phi_{\mathrm{opt}} \times \frac{d^{2}}{A} &= \frac{1}{4\pi^{2}}
    \ln\left(1 + \frac{\zeta_{\mathrm{A}}
      \zeta_{\mathrm{B}}}{4}\right)\Theta(4 - \zeta_{\mathrm{A}}
    \zeta_{\mathrm{B}}) \nonumber \\ &\hspace{-0.6in}+
    \frac{1}{8\pi^{2}}\left[\ln(\zeta_{\mathrm{A}} \zeta_{\mathrm{B}})
      + \ln^2\left(\frac{\zeta_{\mathrm{A}}
        \zeta_{\mathrm{B}}}{4}\right)^{1/2}\right]\Theta(\zeta_{\mathrm{A}}
    \zeta_{\mathrm{B}} - 4), %% \\ \Phi_{\mathrm{sc}} \times
    %% \frac{d^{2}}{A} &= \frac{1}{4\pi^{2}} \ln\left(1 +
    %% \frac{\zeta_{\mathrm{A}} \zeta_{\mathrm{B}}}{4}\right)
    %% \\ \Phi_{\mathrm{Born}} \times \frac{d^{2}}{A} &=
    %% \frac{\zeta_{\mathrm{A}} \zeta_{\mathrm{B}}}{16\pi^{2}}.
%  \end{split}
\end{align}
while $\Phi_{\mathrm{sc}} \times \frac{d^{2}}{A}$ is given by the
first term in \eqref{Phiopt_ext} (without the Heaviside step function)
and $\Phi_{\mathrm{Born}} \times \frac{d^{2}}{A} =
\frac{\zeta_{\mathrm{A}} \zeta_{\mathrm{B}}}{16\pi^{2}}$. As observed
in \figref{platebounds}(a), all three bounds converge to one another
for small $\zeta_{\mathrm{A}} \zeta_{\mathrm{B}}$, with
$\Phi_{\mathrm{opt}} = \Phi_{\mathrm{sc}}$ for $\zeta_{\mathrm{A}}
\zeta_{\mathrm{B}} \leq 4$. As $\zeta_{\mathrm{A}} \zeta_{\mathrm{B}}$
increases, the Born limits grossly overestimate the extent to which
NFRHT can be optimized due to its simple linear dependence on
$\zeta_{\mathrm{A}} \zeta_{\mathrm{B}}$, whereas the exact bound and
scalar approximation grow with respect to $\zeta_{\mathrm{A}}
\zeta_{\mathrm{B}}$ in a much slower logarithmic fashion. Strictly
speaking, $\Phi_{\mathrm{opt}}$ grows faster than $\Phi_{\mathrm{sc}}$
as the latter grows as a logarithm while the former grows as the
square of a logarithm, but in practice the difference is minute:
$\sqrt{\zeta_{\mathrm{A}} \zeta_{\mathrm{B}}}$ would have to reach
$10^{6}$ for the two quantities to differ even by a factor of 4. As a
consequence, the bound can practically be reached by homogeneous
isotropic planar bodies at the surface polariton resonance condition
$\Re(1/\chi) = -1/2$, and the enhancement of $\Phi \times
\frac{d^{2}}{A}$ relative to $\frac{1}{4\pi^{2}}$ will be $O(1)$ at
best in practice regardless of the actual value of $\chi$ there. Thus,
even more so than for dipolar bodies, there is very little room for
improving $\Phi$ through nanostructuring compared to what can be
achieved by planar polar-dielectric films.

We also evaluate $\Phi_{\mathrm{opt}}$ and $\Phi_{\mathrm{Born}}$ for
planar films of finite thickness $h$ [\figref{platebounds}(b)], and
point out that each of these bounds only depends on $d$ and $h$ via
the common term $\frac{A}{d^{2}}$ and via a function that depends only
on $\zeta_{\mathrm{A}} \zeta_{\mathrm{B}}$ and the ratio $h/d$. In
particular, we find that for thin films (compared to the separation),
$\Phi_{\mathrm{opt}}$ converges to $\Phi_{\mathrm{Born}}$ for
decreasing thickness at each value of $\zeta =
\sqrt{\zeta_{\mathrm{A}} \zeta_{\mathrm{B}}}$, consistent with
decreasing multiple scattering.  However, as the thickness increases
even to $h/d \approx 0.1$, each of these bounds quickly approaches its
respective bulk asymptote in the limit $h/d \to \infty$. Moreover, the
logarithmic scale on the plot makes clear that these asymptotic values
of $\Phi_{\mathrm{Born}}$ grow linearly with $\zeta_{\mathrm{A}}
\zeta_{\mathrm{B}}$, whereas the corresponding growth of
$\Phi_{\mathrm{opt}}$ is logarithmic.
% with respect to $\zeta_{\mathrm{A}} \zeta_{\mathrm{B}}$.
We do not show $\Phi_{\mathrm{sc}}$ because it is so close to
$\Phi_{\mathrm{opt}}$ in these regimes that the curves would
%overlap and would 
be difficult to distinguish; this again suggests that while reaching
the exact bounds for a given thickness $h$ would require nanoscale
texturing, the bounds can be practically reached by planar films of
the same thickness and appropriately chosen materials, in line with
previous observations restricted to one-dimensionally periodic
media~\cite{MillerPRL2014}.

\begin{figure}[t!]
\centering
\includegraphics[width=0.9\columnwidth]{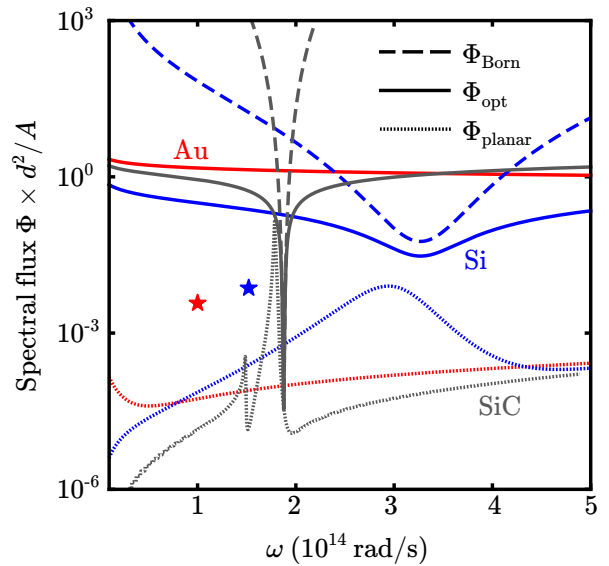}
\caption{Comparison of $\Phi_{\mathrm{Born}}$ (dashed) and
  $\Phi_{\mathrm{opt}}$ (solid) for extended bodies to planar heat
  transfer $\Phi_{\mathrm{planar}}$ (dotted) at frequencies relevant
  to the Planck function at typical experimental temperatures,
  considering Au (red), doped Si (blue), and SiC (dark gray). Also
  shown are the maximum $\Phi$ of representative nanostructured Au
  (red star)~\cite{MessinaPRB2017} and doped Si (blue
  star)~\cite{FernandezHurtadoPRL2017}
  surfaces. $\Phi_{\mathrm{Born}}$ for Au is several orders of
  magnitude above the plotted range and thus not shown.}
\label{fig:platereal}
\end{figure}

Finally, we compare the power spectrum $\Phi_{\mathrm{planar}} \times
d^{2} / A$ associated with identical planar films~\cite{MillerPRL2015,
  JinOE17} to the exact and Born bounds in~\figref{platereal},
specifically considering gold (Au), doped silicon (Si), and silicon
carbide (SiC) as representative materials, as well as to the largest
heat transfer observed in specific nanostructured
Au~\cite{MessinaPRB2017} and Si~\cite{FernandezHurtadoPRL2017}
surfaces studied in the past. (We employ Drude dispersions for
Au~\cite{MessinaPRB2017} and Si~\cite{FernandezHurtadoPRL2017}, and a
phonon polaritonic dispersion for SiC~\cite{HongMRE2018}.) In
particular, in the infrared where the Planck function is considerable
(at typical experimental temperatures, $T \lesssim 1000~\mathrm{K}$),
$\Phi_{\mathrm{Born}}$ for all of these materials is significantly
larger than the corresponding $\Phi_{\mathrm{opt}}$ and is highly
sensitive to material dispersion; as a specific example, the Born
bound for Au lies significantly above the upper limits of the plot
over the entire range of frequencies shown. By contrast, the
logarithmic dependence of $\Phi_{\mathrm{opt}}$ on $\zeta_p$ means
that it will generally be much less sensitive to changes in material
dispersion except near polariton resonances; this is noticeable in the
infrared for Si and more so for SiC, whereas Au does not feature
material resonances except at much higher frequencies.  We find that
$\Phi_\mathrm{planar}$ is consistently much smaller than either
$\Phi_{\mathrm{Born}}$ or $\Phi_{\mathrm{opt}}$ for Au owing to the
lack of infrared resonances; the Au nanostructures of
\cite{MessinaPRB2017} improve on the results for Au plates by two
orders of magnitude, but still fall more than two orders of magnitude
shy of $\Phi_{\mathrm{opt}}$ at that frequency. The outlook is more
pessimistic for polar dielectrics like doped Si or
SiC. Nanostructuring Si into a metasurface as in
\cite{FernandezHurtadoPRL2017} barely improves $\Phi$ above the peak
of the planar result, which never reaches its bound because the
dispersion of Si prohibits the planar surface plasmon resonance
condition $\Re(1/\chi) = -1/2$ from being reached; only the integrated
NFRHT power $P$ increases substantially by virtue of the peak
frequency being much smaller (i.e. escaping the exponential
suppression of the Planck function). Meanwhile, SiC plates exhibit a
power spectrum $\Phi$ that touches $\Phi_{\mathrm{opt}}$ at two
points, the smaller of which is the material resonance where the
losses become so large that the exact and Born limits coincide (as we
have that shown multiple scattering becomes irrelevant for large
dissipation), and the larger of which is a polaritonic resonance where
$\Phi_{\mathrm{opt}}$ is nearly constant while $\Phib$ is larger by an
unattainable factor of 50; we note that at those resonances,
$\Phi_{\mathrm{planar}} = \Phi_{\mathrm{sc}}$.

\emph{Concluding remarks.---} The results above suggest that apart
from redshifting resonance frequencies to improve $P$ (especially
useful for metals), nanostructuring of either dipolar or extended
media cannot produce significantly better results for $\Phi$ than do
spherical or planar objects, eventually saturating or exhibiting a
logarithmic dependence on $\zeta = |\chi|^2/\Im(\chi)$ in each case.
At first glance, this is a surprising contrast to the success of
nanostructuring in enhancing the local density of
states~\cite{MillerOE2016}.  This dichotomy can be understood as a
consequence of finite-size effects: a dipole radiator does not scatter
fields and hence an infinite number of modes can participate in
absorption, but this cannot hold for objects of finite size.

While we have focused on NFRHT at individual resonance frequencies,
their narrow bandwidths $\Delta\omega \sim \omega
\frac{\Im(\chi)}{|\chi|}$ permit approximate bounds on the integrated
heat transfer~\cite{MillerPRL2015}. For two bodies of the same
susceptibility $\chi$, this yields:
\begin{equation*}
  P_\mathrm{opt} \approx \frac{\omega \Im(\chi)}{|\chi|}
  \Phi_{\mathrm{opt}}(\omega) [\Pi(\omega, T_{\mathrm{B}}) -
    \Pi(\omega, T_{\mathrm{A}})].
\end{equation*}
For dipolar bodies, $\Phi_{\mathrm{opt}}$ reaches a maximum with
respect to $\zeta$ and never diverges, while for extended structures
the divergence is logarithmic. Hence, beyond a threshold, any increase
in $\Phi_{\mathrm{opt}}$ from larger material response will be
accompanied by a corresponding decrease in $\Delta\omega$; this
suggests that regardless of object sizes, there exists an optimal
$\zeta$ maximizing $P$.

Finally, we emphasize that the above analyses focused on the
near-field, which can be justified for small enough separations, but
$\Phi_{\mathrm{opt}}$ and $\Phi_{\mathrm{sc}}$ in general can be
evaluated at every lengthscale, whereas the same cannot be said of
$\Phi_{\mathrm{Born}}$. That said, as discussed in
\cite{MoleskyARXIV2019A}, our bounds do not explicitly include the
effects of far-field radiative losses, which in conjunction with
multiple scattering should provide even tighter bounds. Additionally,
similar bounds could be derived for other problems in fluctuational
electromagnetism, including fluorescence energy
transfer~\cite{PolimeridisPRB2015} and Casimir
forces~\cite{KennethPRL2006}, the subject of future work.

\emph{Acknowledgments.---}The authors would like to thank Riccardo
Messina and Pengning Chao for helpful discussions. This work was
supported by the National Science Foundation under Grants
No. DMR-1454836, DMR 1420541, DGE 1148900, the Cornell Center for
Materials Research MRSEC (award no. DMR-1719875), and the Defense
Advanced Research Projects Agency (DARPA) under agreement
HR00111820046. The views, opinions and/or findings expressed are those
of the authors and should not be interpreted as representing the
official views or policies of the Department of Defense or the
U.S. Government.

\appendix

\section{Notation}

We briefly discuss the notation used through the main text and the
appendices. A vector field $\vec{v}(\vec{x})$ will be denoted as
$\ket{\vec{v}}$. The conjugated inner product is $\bracket{\vec{u},
  \vec{v}} = \int~\mathrm{d}^{3} x~\vec{u}^{\star} (\vec{x}) \cdot
\vec{v}(\vec{x})$. An operator $\mathbb{A}(\vec{x}, \vec{x}')$ will be
denoted as $\mathbb{A}$, with $\int~\mathrm{d}^{3}
x'~\mathbb{A}(\vec{x}, \vec{x}') \cdot \vec{v}(\vec{x}')$ denoted as
$\mathbb{A}\ket{\vec{v}}$. The Hermitian conjugate
$\mathbb{A}^{\dagger}$ is defined such that $\bracket{\vec{u},
  \mathbb{A}^{\dagger} \vec{v}} = \bracket{\mathbb{A} \vec{u},
  \vec{v}}$. The anti-Hermitian part of a square operator (whose
domain and range are the same size) is defined as the operator
$\asym(\mathbb{A}) = (\mathbb{A} -
\mathbb{A}^{\dagger})/(2\im)$. Finally, the trace of an operator is
$\tr(\mathbb{A}) = \int~\mathrm{d}^{3} x~\tr(\mathbb{A}(\vec{x},
\vec{x}))$. Through this paper, unless stated explicitly otherwise,
all quantities implicitly depend on $\omega$, and such dependence will
be notationally suppressed for brevity.

\section{Properties of $\Phi_{\mathrm{sc}}$}

In this section, we show that the scalar approximation to the bound on
NFRHT between two bodies A and B in vacuum exhibits a local stationary
point when both bodies satisfy the optimal absorption condition in
isolation. We also show that the scalar approximation in the
near-field is domain monotonic, meaning that it can be evaluated for
larger domains than the bodies in question given their material
response factors. These results make use of the fact that in the
absence of retardation, $\Gvac_{\mathrm{BA}} =
(\Gvac_{\mathrm{AB}})^{\top}$ is a real-valued operator in
position-space, so $\Gvac_{\mathrm{BA}} \Gvac_{\mathrm{AB}}$ is a
Hermitian positive-semidefinite operator.

\subsection{Stationarity of the scalar approximation}

In this section, we prove that $\Phi_{\mathrm{sc}}$ in the near-field
exhibits a local stationary point when the
T-operators~\cite{MoleskyARXIV2019A} of each body satisfy the
condition of zero far-field scattering in isolation. Thus, if body A
is fixed to be an isolated perfect absorber satisfying
$\TT_{\mathrm{A}} = \im\zeta_{\mathrm{A}} \II_{\mathrm{A}}$, then any
change to body B from perfect absorption, written as $\TT_{\mathrm{B}}
= \zeta_{\mathrm{B}} \left(\im\II_{\mathrm{B}} +
\zeta_{\mathrm{B}}^{-1} \mathbb{R}\right)$ for a small perturbation
$\mathbb{R}$ (restricted to be real symmetric to preserve the
condition of zero far-field scattering by $\TT_{\mathrm{B}}$),
produces no change in the NFRHT to first order. By reciprocity, the
same arguments hold if A and B are exchanged.

Defining the real symmetric positive-semidefinite operator $\mathbb{K}
= \zeta_{\mathrm{A}} \zeta_{\mathrm{B}} \Gvac_{\mathrm{BA}}
\Gvac_{\mathrm{AB}}$ and replacing $\II_{\mathrm{B}}$ by $\II$ for
notational convenience, NFRHT may be written as
\begin{multline}
  \Phi = \frac{2}{\pi} \tr\Bigg[\left(\II + \mathbb{K} +
    i\mathbb{K}\mathbb{R}\right)^{-1} \times \\ \left(\II_{\mathrm{B}}
    - (-i\II + \mathbb{R}) \frac{\Im(\Gvac)}{\lambda_{\mathrm{B}}}
    (i\II + \mathbb{R})\right) \times \\ \left(\II + \mathbb{K} -
    i\mathbb{K}\mathbb{R}\right)^{-1} \mathbb{K}\Bigg]
\end{multline}
where we have used the facts that
$\TT_{\mathrm{A}} \Im(\VV_{\mathrm{A}}^{-1\star})
\TT_{\mathrm{A}}^{\star} = \zeta_\mathrm{A}
\II_{\mathrm{A}}$ and that in general,
$\TT_{\mathrm{B}}^{\star} \Im(\VV_{\mathrm{B}}^{-1\star})
\TT_{\mathrm{B}} = \Im(\TT_{\mathrm{B}}) - \TT_{\mathrm{B}}^{\star}
\Im(\Gvac) \TT_{\mathrm{B}}$, after which point the definition of
$\TT_{\mathrm{B}}$ in terms of $\mathbb{R}$ may be substituted. This
trace can be expanded order-by-order in $\mathbb{R}$, with
$\Phi^{(n)}$ denoting the $n$th order term.

The lowest-order term is given by,
\begin{equation}
  \Phi_{\mathrm{sc}}^{(0)} = \frac{2}{\pi} \tr\Bigg[\left(\II +
    \mathbb{K}\right)^{-1} \left(\II_{\mathrm{B}} -
    \frac{\Im(\Gvac)}{\lambda_{\mathrm{B}}}\right)\left(\II +
    \mathbb{K}\right)^{-1} \mathbb{K}\Bigg]
\end{equation}
which, upon undoing the substitution
$\TT_{\mathrm{B}}^{\star} \Im(\VV_{\mathrm{B}}^{-1\star})
\TT_{\mathrm{B}} = \Im(\TT_{\mathrm{B}}) - \TT_{\mathrm{B}}^{\star}
\Im(\Gvac) \TT_{\mathrm{B}}$ and the definition of $\TT_{\mathrm{B}}$
in terms of $\mathbb{R}$, is identical to the result in the main text.

The first-order term is given by,
\begin{multline}
  \Phi_{\mathrm{sc}}^{(1)} = \frac{2}{\pi}
  \tr\Bigg[-\frac{i}{\lambda_{\mathrm{B}}} (\II + \mathbb{K})^{-1}
    \mathbb{R}\Im(\Gvac) (\II + \mathbb{K})^{-1} \mathbb{K} +
    \\ \frac{i}{\lambda_{\mathrm{B}}} (\II + \mathbb{K})^{-1}
    \Im(\Gvac) \mathbb{R} (\II + \mathbb{K})^{-1} \mathbb{K} -
    \\ i(\II + \mathbb{K})^{-1} \mathbb{R} \mathbb{K} (\II +
    \mathbb{K})^{-1} \left(\II -
    \frac{\Im(\Gvac)}{\lambda_{\mathrm{B}}}\right) (\II +
    \mathbb{K})^{-1} \mathbb{K} + \\ i(\II + \mathbb{K})^{-1}
    \left(\II - \frac{\Im(\Gvac)}{\lambda_{\mathrm{B}}}\right) (\II +
    \mathbb{K})^{-1} \mathbb{K} \mathbb{R} (\II + \mathbb{K})^{-1}
    \mathbb{K} \Bigg]
\end{multline}
but by exploiting the invariance of the trace under cyclic permutation
and transposition, and noting that $\mathbb{K} = \mathbb{K}^{\top}$
and $\mathbb{R} = \mathbb{R}^{\top}$, this trace actually
vanishes. Therefore, each body satisfying perfect absorption in
isolation produces a local stationary point in $\Phi_{\mathrm{sc}}$.

\subsection{Domain monotonicity of $\Phi_{\mathrm{sc}}$}

We now prove that the $\Phi_{\mathrm{sc}}$ factor is domain monotonic,
meaning that it will always increase when the spatial domain (i.e. the
volume of either body) increases; this has previously been proven for
the scalar Laplace operator with Dirichlet
boundaries~\cite{GrebenkovSIAM2013} but to our knowledge, not for
$\frac{2}{\pi} \zeta_{\mathrm{A}} \zeta_{\mathrm{B}} \left\lVert
(\II_{\mathrm{B}} + \zeta_{\mathrm{A}} \zeta_{\mathrm{B}}
\Gvac_{\mathrm{BA}} \Gvac_{\mathrm{AB}})^{-1} \Gvac_{\mathrm{BA}}
\right\rVert_{\mathrm{F}}^{2}$. We allow bodies A and B to have
different shapes, sizes, and material response factors $\zeta_{p}$ for
$p \in \{\mathrm{A}, \mathrm{B}\}$, and we assume only that
$\zeta_{p}$ as well as the minimum separation $d$ are fixed throughout
this proof. In particular, we assume a small enough perturbative
increase to the volume of either object so that each object remains an
optimal absorber even with the new volume, i.e. $\TT_{p} =
\im\zeta_{p} \II_{p}$ is still true even with the new degrees of
freedom. If body B undergoes a perturbative increase in volume while
body A remains unchanged, the projection operator onto the original
volume of B (comprising the actual material degrees of freedom, not
the entire convex hull, which is relevant if the original volume of B
has interior holes or surface concavities) will be denoted as
$\mathbb{P}_{0}$, while the projection operator onto the added
material volume in B will be denoted as $\mathbb{P}_{\Delta}$, with
$\mathbb{P}_{0} \mathbb{P}_{\Delta} = \mathbb{P}_{\Delta}
\mathbb{P}_{0} = 0$ encoding the disjointness of the two
spaces. Denoting $\Gvac_{\mathrm{B}_{0} \mathrm{A}} = \mathbb{P}_{0}
\Gvac_{\mathrm{BA}}$, $\Gvac_{\Delta\mathrm{BA}} = \mathbb{P}_{\Delta}
\Gvac_{\mathrm{BA}}$, $\Gvac_{\mathrm{AB}_{0}} =
(\Gvac_{\mathrm{B}_{0} \mathrm{A}})^{\top}$, and $\Gvac_{\mathrm{A}
  \Delta\mathrm{B}} = (\Gvac_{\Delta\mathrm{BA}})^{\top}$, and
defining
\begin{equation}
  \begin{split}
    \Gvac_{\mathrm{BA}} &= \begin{bmatrix}
      \Gvac_{\mathrm{B}_{0} \mathrm{A}} \\
      \Gvac_{\Delta\mathrm{BA}}
    \end{bmatrix} \\
    \Gvac_{\mathrm{AB}} &= \begin{bmatrix}
      \Gvac_{\mathrm{AB}_{0}} &
      \Gvac_{\mathrm{A} \Delta\mathrm{B}}
    \end{bmatrix} \\
    \II_{\mathrm{B}} &= \begin{bmatrix}
      \mathbb{P}_{0} & 0 \\
      0 & \mathbb{P}_{\Delta}
    \end{bmatrix}
  \end{split}
\end{equation}
allows for writing (in a slight abuse of notation)
\begin{equation}
  \begin{split}
    \Gvac_{\mathrm{BA}} \Gvac_{\mathrm{AB}} &\equiv \Gvac_{\mathrm{B}_{0} \mathrm{A}} \Gvac_{\mathrm{AB}_{0}} + \Delta(\Gvac_{\mathrm{BA}} \Gvac_{\mathrm{AB}}) \\
    \Gvac_{\mathrm{B}_{0} \mathrm{A}}
    \Gvac_{\mathrm{AB}_{0}}
    &\equiv \begin{bmatrix}
      \Gvac_{\mathrm{B}_{0} \mathrm{A}} \Gvac_{\mathrm{AB}_{0}} & 0 \\
      0 & 0
    \end{bmatrix} \\
    \Delta(\Gvac_{\mathrm{BA}}
    \Gvac_{\mathrm{AB}}) &\equiv \begin{bmatrix}
      0 & \Gvac_{\mathrm{B}_{0} \mathrm{A}} \Gvac_{\mathrm{A}\Delta\mathrm{B}} \\
      \Gvac_{\Delta\mathrm{BA}}
      \Gvac_{\mathrm{AB}_{0}} &
      \Gvac_{\Delta\mathrm{BA}}
      \Gvac_{\mathrm{A}\Delta\mathrm{B}}
    \end{bmatrix}
  \end{split}
\end{equation}
for this system. This in turn leads to the expression,
\begin{multline}
  (\II_{\mathrm{B}} + \zeta_{\mathrm{A}} \zeta_{\mathrm{B}}
  \Gvac_{\mathrm{BA}} \Gvac_{\mathrm{AB}})^{-1} 
\\ = (\II_{\mathrm{B}} +
  \zeta_{\mathrm{A}} \zeta_{\mathrm{B}}
  \Gvac_{\mathrm{B}_{0} \mathrm{A}} \Gvac_{\mathrm{AB}_{0}})^{-1} 
  \\ - \zeta_{\mathrm{A}} \zeta_{\mathrm{B}}
  (\II_{\mathrm{B}} + \zeta_{\mathrm{A}} \zeta_{\mathrm{B}}
  \Gvac_{\mathrm{B}_{0} \mathrm{A}} \Gvac_{\mathrm{AB}_{0}})^{-1}
\\ \times
\Delta(\Gvac_{\mathrm{BA}} \Gvac_{\mathrm{AB}})(\II_{\mathrm{B}} +
  \zeta_{\mathrm{A}} \zeta_{\mathrm{B}}
  \Gvac_{\mathrm{B}_{0} \mathrm{A}} \Gvac_{\mathrm{AB}_{0}})^{-1} 
\\ 
+  O((\Delta(\Gvac_{\mathrm{BA}} \Gvac_{\mathrm{AB}}))^{2}),
\end{multline}
to lowest order in the term $\Delta(\Gvac_{\mathrm{BA}}
\Gvac_{\mathrm{AB}})$, which is small as the addition to the volume of
B is small (perturbative). Plugging this into the expression for
$\Phi_{\mathrm{sc}}$ and exploiting the cyclic property of the trace
for notational convenience yields,
\begin{multline}
  \tr(\Gvac_{\mathrm{BA}} \Gvac_{\mathrm{AB}} (\II_{\mathrm{B}} +
  \zeta_{\mathrm{A}} \zeta_{\mathrm{B}} \Gvac_{\mathrm{BA}}
  \Gvac_{\mathrm{AB}})^{-2}) = \\ \tr(\Gvac_{\mathrm{B}_{0}
    \mathrm{A}} \Gvac_{\mathrm{AB}_{0}} (\II_{\mathrm{B}} +
  \zeta_{\mathrm{A}} \zeta_{\mathrm{B}} \Gvac_{\mathrm{B}_{0}
    \mathrm{A}} \Gvac_{\mathrm{AB}_{0}})^{-2}) \\ +
  \tr(\Delta(\Gvac_{\mathrm{BA}} \Gvac_{\mathrm{AB}})
  (\II_{\mathrm{B}} + \zeta_{\mathrm{A}} \zeta_{\mathrm{B}}
  \Gvac_{\mathrm{B}_{0} \mathrm{A}} \Gvac_{\mathrm{AB}_{0}})^{-2})
  \\ - 2\zeta_{\mathrm{A}} \zeta_{\mathrm{B}}
  \tr(\Delta(\Gvac_{\mathrm{BA}} \Gvac_{\mathrm{AB}})(\II_{\mathrm{B}}
  + \zeta_{\mathrm{A}} \zeta_{\mathrm{B}} \Gvac_{\mathrm{B}_{0}
    \mathrm{A}} \Gvac_{\mathrm{AB}_{0}})^{-3} \\ \times
  \Gvac_{\mathrm{B}_{0} \mathrm{A}} \Gvac_{\mathrm{AB}_{0}}) +
  O((\Delta(\Gvac_{\mathrm{BA}} \Gvac_{\mathrm{AB}}))^{2})
\end{multline}
to lowest order in the term $\Delta(\Gvac_{\mathrm{BA}}
\Gvac_{\mathrm{AB}})$, for which each of the three terms may be
analyzed individually. The first term is merely the unperturbed
contribution to $\Phi_{\mathrm{sc}}$, so the perturbation to lowest
order comprises the second and third terms. For the second term, the
factor
%\begin{equation*}
\begin{multline*}
  (\II_{\mathrm{B}} + \zeta_{\mathrm{A}} \zeta_{\mathrm{B}}
  \Gvac_{\mathrm{B}_{0} \mathrm{A}} \Gvac_{\mathrm{AB}_{0}})^{-2}
  \\ = \begin{bmatrix} (\mathbb{P}_{0} + \zeta_{\mathrm{A}}
    \zeta_{\mathrm{B}} \Gvac_{\mathrm{B}_{0} \mathrm{A}}
    \Gvac_{\mathrm{AB}_{0}})^{-2} & 0 \\ 0 & \mathbb{P}_{\Delta}
    \end{bmatrix}
\end{multline*}
%\end{equation*}
leads to
%\begin{equation*}
\begin{multline*}
%  \begin{split}
    \Delta(\Gvac_{\mathrm{BA}} \Gvac_{\mathrm{AB}})(\II_{\mathrm{B}} +
    \zeta_{\mathrm{A}} \zeta_{\mathrm{B}} \Gvac_{\mathrm{B}_{0}
      \mathrm{A}} \Gvac_{\mathrm{AB}_{0}})^{-2} \\ = \begin{bmatrix} 0
      & \Gvac_{\mathrm{B}_{0} \mathrm{A}}
      \Gvac_{\mathrm{A}\Delta\mathrm{B}} \\ \Gvac_{\Delta\mathrm{BA}}
      \Gvac_{\mathrm{AB}_{0}} (\mathbb{P}_{0} + \zeta_{\mathrm{A}}
      \zeta_{\mathrm{B}} \Gvac_{\mathrm{B}_{0} \mathrm{A}}
      \Gvac_{\mathrm{AB}_{0}})^{-2} & \Gvac_{\Delta\mathrm{BA}}
      \Gvac_{\mathrm{A}\Delta\mathrm{B}}
    \end{bmatrix}
\end{multline*}
%\end{equation*}
whose trace is simply
$\tr(\Gvac_{\Delta\mathrm{BA}}
\Gvac_{\mathrm{A}\Delta\mathrm{B}})$. For the third term, the factor
\begin{multline*}
%\begin{equation*}
  (\II_{\mathrm{B}} + \zeta_{\mathrm{A}} \zeta_{\mathrm{B}}
  \Gvac_{\mathrm{B}_{0} \mathrm{A}} \Gvac_{\mathrm{AB}_{0}})^{-3}
  \Gvac_{\mathrm{B}_{0} \mathrm{A}} \Gvac_{\mathrm{AB}_{0}}
  \\ = \begin{bmatrix} (\mathbb{P}_{0} + \zeta_{\mathrm{A}}
    \zeta_{\mathrm{B}} \Gvac_{\mathrm{B}_{0} \mathrm{A}}
    \Gvac_{\mathrm{AB}_{0}})^{-3} \Gvac_{\mathrm{B}_{0} \mathrm{A}}
    \Gvac_{\mathrm{AB}_{0}} & 0 \\ 0 & 0
    \end{bmatrix}
%\end{equation*}
\end{multline*}
leads to
\begin{equation*}
  \begin{split}
    &\Delta(\Gvac_{\mathrm{BA}} \Gvac_{\mathrm{AB}})(\II_{\mathrm{B}}
    + \zeta_{\mathrm{A}} \zeta_{\mathrm{B}} \Gvac_{\mathrm{B}_{0}
      \mathrm{A}} \Gvac_{\mathrm{AB}_{0}})^{-3} \Gvac_{\mathrm{B}_{0}
      \mathrm{A}} \Gvac_{\mathrm{AB}_{0}} = \\ &\begin{bmatrix} 0 & 0
      \\ \Gvac_{\Delta\mathrm{BA}} \Gvac_{\mathrm{AB}_{0}}
      (\mathbb{P}_{0} + \zeta_{\mathrm{A}} \zeta_{\mathrm{B}}
      \Gvac_{\mathrm{B}_{0} \mathrm{A}} \Gvac_{\mathrm{AB}_{0}})^{-3}
      \Gvac_{\mathrm{B}_{0} \mathrm{A}} \Gvac_{\mathrm{AB}_{0}} & 0
    \end{bmatrix}
  \end{split}
\end{equation*}
whose trace vanishes. Therefore, a perturbative increase in the volume
of body B changes the contribution to $\Phi_{\mathrm{sc}}$ by an
amount $\tr(\Gvac_{\Delta\mathrm{BA}}
\Gvac_{\mathrm{A}\Delta\mathrm{B}})$, independent of $\zeta_{p}$ for
$p \in \{\mathrm{A}, \mathrm{B}\}$; as $\Gvac_{\Delta\mathrm{BA}} =
(\Gvac_{\mathrm{A}\Delta\mathrm{B}})^{\top}$ is real-valued in the
near-field, then $\Gvac_{\Delta\mathrm{BA}}
\Gvac_{\mathrm{A}\Delta\mathrm{B}}$ is real-symmetric
positive-semidefinite, so its trace is nonnegative, and is exactly the
pairwise additive contribution to $\left\lVert \Gvac_{\mathrm{BA}}
\right\rVert_{\mathrm{F}}^{2}$ (in the absence of multiple scattering)
from the same perturbation. Reciprocity implies invariance of this
contribution to $\Phi_{\mathrm{sc}}$ under interchange of bodies A and
B, which means that the same arguments can be used to show that a
perturbative increase in the volume of A (holding B fixed) increases
the contribution to $\Phi_{\mathrm{sc}}$. As both of these statements
are true regardless of the original geometries of A and B, they must
remain true for any combination of increases in the volumes of A and
B, even if the minimum separation $d$ does not change. As a result,
for a given $d$ and $\zeta_{p}$ for $p \in \{\mathrm{A},
\mathrm{B}\}$, the volume that maximizes the domain of the scattering
operators (a planar semi-infinite half-space and its geometric mirror
image, though $\zeta_{\mathrm{A}}$ and $\zeta_{\mathrm{B}}$ may
differ), leads to their largest $\Phi_{\mathrm{sc}}$. For such
restricted T-operators, nanostructuring will therefore always decrease
$\Phi_{\mathrm{sc}}$ for fixed $d$ and material response factors.

\section{Singular values of $\Gvac_{\mathrm{BA}}$ for dipolar particles}

In this section, we derive analytical expressions for the singular
values $g_{i}$ of $\Gvac_{\mathrm{BA}}$ in the near-field, where body
B is a dipolar nanoparticle and body A is either another dipolar
nanoparticle or an extended object. We start with the case of two
dipoles. This means for each body $p \in \{\mathrm{A}, \mathrm{B}\}$,
the relevant basis functions are $\vec{a}_{i} (\vec{x}) =
\sqrt{V_{\mathrm{A}}} \delta^{3} (\vec{x} - \vec{r}_{\mathrm{A}})
\vec{e}_{i}$ and $\vec{b}_{i} (\vec{x}) = \sqrt{V_{\mathrm{B}}}
\delta^{3} (\vec{x} - \vec{r}_{\mathrm{B}}) \vec{e}_{i}$. Without loss
of generality, we take $\vec{r}_{\mathrm{A}} = 0$ and
$\vec{r}_{\mathrm{B}} = d\vec{e}_{z}$. This means that we write the
near-field Green's function tensor in position space as
$\bracket{\vec{b}_{i}, \Gvac_{\mathrm{BA}} \vec{a}_{j}} =
\frac{\sqrt{V_{\mathrm{A}} V_{\mathrm{B}}}}{4\pi d^{3}}
\left(3\delta_{i, 3} \delta_{j,3} - \delta_{ij}\right)$. As a result,
we may immediately read off the singular values $g_{1} = g_{2} =
\frac{\sqrt{V_{\mathrm{A}} V_{\mathrm{B}}}}{4\pi d^{3}}$ and $g_{3} =
2g_{1} = \frac{\sqrt{V_{\mathrm{A}} V_{\mathrm{B}}}}{2\pi d^{3}}$.

We now consider a situation in which body B remains dipolar but body A
is replaced by an extended object enclosed by the semi-infinite
half-space $z \leq 0$; for simplicity, we will denote $V_{\mathrm{B}}$
simply as $V$. Without loss of generality, we still take
$\vec{r}_{\mathrm{B}} = d\vec{e}_{z}$ and $\vec{b}_{i} (\vec{x}) =
\sqrt{V_{\mathrm{B}}} \delta^{3} (\vec{x} - \vec{r}_{\mathrm{B}})
\vec{e}_{i}$.  Normalizable basis functions for body A are harder to
define due to the semi-infinite domain.  However, because the singular
values of $\Gvac_{\mathrm{BA}}$ are simply the eigenvalues of
$\Gvac_{\mathrm{BA}} \GG_{\mathrm{AB}}^{\mathrm{vac}\star}$, and
because $\Gvac_{\mathrm{BA}}$ is real-valued in the near-field, we
need only to evaluate the matrix elements $\bracket{\vec{b}_{i},
  \Gvac_{\mathrm{BA}} \Gvac_{\mathrm{AB}} \vec{b}_{j}}$, where the
operator product $\Gvac_{\mathrm{BA}} \Gvac_{\mathrm{AB}}$ can be
evaluated in position space. This evaluation yields
\begin{multline*}
  \sum_{k} \int_{V_{\mathrm{A}}} G^{\mathrm{vac}}_{ik}
  (\vec{r}_{\mathrm{B}}, \vec{x}) \cdot G^{\mathrm{vac}}_{kj}
  (\vec{x}, \vec{r}_{\mathrm{B}})~\mathrm{d}^{3} \vec{x} = \\V
  \int_{-\infty}^{\infty} \int_{-\infty}^{\infty} \int_{-\infty}^{0}
  \frac{\mathrm{d}x~\mathrm{d}y~\mathrm{d}z}{16\pi^{2}
    |\vec{r}_{\mathrm{B}} - \vec{x}|^{6}} \left(\frac{3(d\delta_{i,3}
    - x_{i})(d\delta_{j,3} - x_{j})}{|\vec{r}_{\mathrm{B}} -
    \vec{x}|^{2}} + \delta_{ij}\right)
\end{multline*}
and this integral can be evaluated in cylindrical coordinates with
$\vec{x} = \rho (\cos(\varphi) \vec{e}_{x} + \sin(\varphi)
\vec{e}_{y}) + z\vec{e}_{z}$, so $|\vec{r}_{\mathrm{B}} - \vec{x}|^{2}
= \rho^{2} + (d - z)^{2}$. The term involving $\delta_{ij}$ can easily
be evaluated due to independence from $\varphi$, yielding:
\begin{equation*}
  \frac{V \delta_{ij}}{8\pi} \int_{0}^{\infty} \int_{-\infty}^{0}
  \frac{1}{(\rho^{2} + (d - z)^{2})^{3}}
  \rho~\mathrm{d}z~\mathrm{d}\rho = \frac{V \delta_{ij}}{96\pi d^{3}}
\end{equation*}
by integrating over $\rho$ and then $z$. The term involving
$(d\delta_{i,3} - x_{i})(d\delta_{j,3} - x_{j})$
requires evaluation of this outer product of vectors. In
cylindrical coordinates, this evaluates as the tensor
\begin{multline*}
  (\vec{r}_{\mathrm{B}} - \vec{x}) \otimes (\vec{r}_{\mathrm{B}} -
  \vec{x}) = \\ \begin{bmatrix} \rho^{2} \cos^{2} (\varphi) & \rho^{2}
    \cos(\varphi) \sin(\varphi) & \rho \cos(\varphi) (d - z)
    \\ \rho^{2} \cos(\varphi) \sin(\varphi) & \rho^{2} \sin^{2}
    (\varphi) & \rho \sin(\varphi) (d - z) \\ \rho \cos(\varphi) (d -
    z) & \rho \sin(\varphi) (d - z) & (d - z)^{2}
    \end{bmatrix}
\end{multline*}
for which integration over $\varphi$ makes the off-diagonal elements
vanish, while integration over the diagonal elements gives $\rho^{2}
\int_{0}^{2\pi} \cos^{2} (\varphi)~\mathrm{d}\varphi = \rho^{2}
\int_{0}^{2\pi} \sin^{2} (\varphi)~\mathrm{d}\varphi = \pi \rho^{2}$
for the $xx$- and $yy$-components or $(d - z)^{2} \int_{0}^{2\pi}
\mathrm{d}\varphi = 2\pi (d - z)^{2}$ for the $zz$-component. The
integral over the $xx$- and $yy$-components therefore yield:
\begin{equation*}
  \frac{3V}{16\pi} \int_{0}^{\infty} \int_{-\infty}^{0}
  \frac{\rho^{3}~\mathrm{d}z~\mathrm{d}\rho}{(\rho^{2} + (d -
    z)^{2})^{4}} = \frac{V}{192\pi d^{3}}
\end{equation*}
while the integral over the $zz$-component yields
\begin{equation*}
  \frac{3V}{8\pi} \int_{0}^{\infty} \int_{-\infty}^{0} \frac{\rho (d -
    z)^{2}~\mathrm{d}z~\mathrm{d}\rho}{(\rho^{2} + (d - z)^{2})^{4}}
  =\frac{V}{48\pi d^{3}}.
\end{equation*}
Adding these contributions to the contributions from the prefactor of
$\delta_{ij}$ yields:
\begin{equation}
  \bracket{\vec{b}_{i}, \Gvac_{\mathrm{BA}} \Gvac_{\mathrm{AB}}
    \vec{b}_{j}} = \frac{V}{64\pi d^{3}}(\delta_{ij} + \delta_{i,3}
  \delta_{j,3})
\end{equation}
from which it follows that the singular values are $g_{1} = g_{2} =
\sqrt{\frac{V}{64\pi d^{3}}}$ and $g_{3} = \sqrt{2} g_{1} =
\sqrt{\frac{V}{32\pi d^{3}}}$.

We note that while $\Phi_{\mathrm{opt}}$ is cumbersome to write
analytically due to the presence of Heaviside step functions, it is
relatively easier to write $\Phi_{\mathrm{Born}}$ and
$\Phi_{\mathrm{sc}}$. For two dipolar bodies, we may write
\begin{equation}
  \begin{split}
    \Phi_{\mathrm{Born}} &= \frac{3\zeta_{\mathrm{A}}
      \zeta_{\mathrm{B}} V_{\mathrm{A}} V_{\mathrm{B}}}{4\pi^{3}
      d^{6}} \\ \Phi_{\mathrm{sc}} &= \frac{\zeta_{\mathrm{A}}
      \zeta_{\mathrm{B}} V_{\mathrm{A}} V_{\mathrm{B}}}{4\pi^{3}
      d^{6}} \left[\frac{1}{\left(1 + \frac{\zeta_{\mathrm{A}}
          \zeta_{\mathrm{B}} V_{\mathrm{A}} V_{\mathrm{B}}}{16\pi^{2}
          d^{6}}\right)^{2}} + \frac{2}{\left(1 +
        \frac{\zeta_{\mathrm{A}} \zeta_{\mathrm{B}} V_{\mathrm{A}}
          V_{\mathrm{B}}}{4\pi^{2} d^{6}}\right)^{2}}\right]
  \end{split}
\end{equation}
while for a dipolar body near an extended structure, we may write
\begin{equation}
  \begin{split}
    \Phi_{\mathrm{Born}} &= \frac{\zeta_{\mathrm{A}}
      \zeta_{\mathrm{B}} V}{8\pi^{2}
      d^{3}} \\ \Phi_{\mathrm{sc}} &= \frac{\zeta_{\mathrm{A}}
      \zeta_{\mathrm{B}} V}{16\pi^{2}
      d^{3}} \left[\frac{1}{\left(1 + \frac{\zeta_{\mathrm{A}}
        \zeta_{\mathrm{B}} V_{\mathrm{B}}}{64\pi
        d^{3}}\right)^{2}} + \frac{1}{\left(1 + \frac{\zeta_{\mathrm{A}}
        \zeta_{\mathrm{B}} V_{\mathrm{B}}}{32\pi
        d^{3}}\right)^{2}}\right].
  \end{split}
\end{equation}

\section{Singular values of $\Gvac_{\mathrm{BA}}$ for extended structures}

In this section, we derive the singular values $g_{i}$ of
$\Gvac_{\mathrm{BA}}$ for two extended structures of infinite
area. Domain monotonicity of our bounds allows us to consider bounding
volumes that are homogeneous in the $xy$-plane, so we will show that
the discrete index $i$ may be replaced by a continuous index
representing the wavevector $\vec{k} = k_{x} \vec{e}_{x} + k_{y}
\vec{e}_{y}$ (i.e. $g_{i} \to g(\vec{k})$).

We first consider two extended (semi-infinite) homogeneous half-spaces
separated by a distance $d$. Without loss of generality, we also
assume the geometry to be mirror-symmetric about $z = 0$, so that the
bulk of bodies A and B are respectively defined for $z < -d/2$ and $z
> d/2$. We further define the mirror flip operator $\OO_{\mathrm{AB}}
= (\OO_{\mathrm{BA}})^{\top} = (\OO_{\mathrm{BA}})^{\dagger} =
(\OO_{\mathrm{BA}})^{-1}$ to be the real-valued unitary operation that
maps a vector field from B to its mirror image in A: reciprocity
implies that $\Gvac_{\mathrm{BA}} \OO_{\mathrm{AB}} =
\OO_{\mathrm{BA}} \Gvac_{\mathrm{AB}}$, so $\Gvac_{\mathrm{AB}} =
\OO_{\mathrm{AB}} \Gvac_{\mathrm{BA}} \OO_{\mathrm{AB}}$. We define
the operator $\DD = \Gvac_{\mathrm{BA}} \OO_{\mathrm{AB}}$, so as $\DD
\DD^{\dagger} = \Gvac_{\mathrm{BA}} (\Gvac_{\mathrm{BA}})^{\dagger}$
by the unitarity of $\OO_{\mathrm{AB}}$, then the singular values of
$\Gvac_{\mathrm{BA}}$ are the same as those of $\DD$.

The mirror symmetry of the problem implies that $\mathbb{D}$ is simply
the negative of the scattering Green's function in the volume of body
B due to a perfect electrically conducting plane coinciding with the
mirror plane, chosen here to be $z = 0$. This allows for immediately
writing
\begin{multline}
  \DD (\vec{k}, \vec{k}', z, z') =
  -\frac{\im\omega^{2}}{2c^{2}} (\mathbb{M}^{\mathrm{s}} +
  \mathbb{M}^{\mathrm{p}}) e^{\im k_{z} (z + z')} 
\\ \times (2\pi)^{2} \delta^{2} (\vec{k} - \vec{k}') \Theta(z - d/2) \Theta(z' -
  d/2)
\end{multline}
in terms of $\vec{k} = k_{x} \vec{e}_{x} + k_{y} \vec{e}_{y}$ and
$k_{z} = \sqrt{\frac{\omega^{2}}{c^{2}} - |\vec{k}|^{2}}$, as well as
the 3-by-3 Cartesian tensors $\mathbb{M}^{\mathrm{s}}$ and
$\mathbb{M}^{\mathrm{p}}$ using the Fresnel reflection coefficients
$r^{\mathrm{s}} = -1$ and $r^{\mathrm{p}} = 1$ for the mirror plane;
the lower boundary at $d/2$ encoded in the Heaviside step functions
$\Theta$ arises from the definitions of the basis functions defining
body B. Using the known expressions for $\mathbb{M}^{\mathrm{s}}$ and
$\mathbb{M}^{\mathrm{p}}$~\cite{Novotny2006}, we work in lowest order
in $\omega/c$, with $|\vec{k}| \gg \omega/c$, so this means that the
contributions from the s-polarization disappear, while those from the
p-polarization do not, which is physically consistent with this
near-field nonretarded (electrostatic) approximation; in particular,
$k_{z} \to \im|\vec{k}|$. This allows for writing
\begin{equation}
  -\frac{\im\omega^{2}}{2c^{2}} \mathbb{M}^{\mathrm{p}} = -\frac{1}{2}
  \begin{bmatrix}
    \frac{k_{x}^{2}}{|\vec{k}|} & \frac{k_{x} k_{y}}{|\vec{k}|} & -\im k_{x} \\
    \frac{k_{x} k_{y}}{|\vec{k}|} & \frac{k_{y}^{2}}{|\vec{k}|} & -\im k_{y} \\
    \im k_{x} & \im k_{y} & |\vec{k}|
  \end{bmatrix}
\end{equation}
for which it can be derived that $-\frac{\im\omega^{2}}{2c^{2}}
\mathbb{M}^{\mathrm{p}}$ has two eigenvalues that are zero and one
eigenvalue that is $-|\vec{k}|$; the corresponding eigenvector
(normalized to 1 under the standard conjugated inner product) for the
latter eigenvalue is $\frac{1}{\sqrt{2}|\vec{k}|} (-\im\vec{k} +
|\vec{k}|\vec{e}_{z})$. Meanwhile, the spatial part $e^{-|\vec{k}|(z +
  z')}$ (having substituted $k_{z} = \im|\vec{k}|$) can be rewritten
as $\frac{e^{-|\vec{k}|d}}{2|\vec{k}|} (\sqrt{2|\vec{k}|}
e^{-|\vec{k}|(z - d/2)})(\sqrt{2|\vec{k}|} e^{-|\vec{k}|(z' - d/2)})$,
which is an outer product of functions in the space of
square-integrable functions on the interval $z \in (d/2, \infty)$,
satisfying the normalization condition $\int_{d/2}^{\infty}
(\sqrt{2|\vec{k}|} e^{-|\vec{k}|(z - d/2)})^{2}~\mathrm{d}z =
1$. Putting all of this together allows for writing $\mathbb{D}$ as a
rank-1 operator:
\begin{multline}
  \DD(\vec{k}, \vec{k}', z, z') = -\frac{e^{-|\vec{k}|d}}{2}
  \vec{v}^{(0)} (\vec{k}, z) \otimes \vec{v}^{(0)\star} (\vec{k}, z')
  \\ \times \Theta(z - d/2) \Theta(z' - d/2) (2\pi)^{2} \delta^{2} (\vec{k} -
  \vec{k}')
\end{multline}
having defined the plane-wave eigenfunctions,
\begin{equation}
  \vec{v}^{(0)} (\vec{k}, z) = \frac{1}{\sqrt{|\vec{k}|}} (-\im\vec{k}
  + |\vec{k}|\vec{e}_{z})e^{-|\vec{k}|(z - d/2)}
\end{equation}
normalized such that $\int_{d/2}^{\infty} \vec{v}^{(0)\star} (\vec{k},
z) \cdot \vec{v}^{(0)} (\vec{k}, z)~\mathrm{d}z = 1$, with
corresponding eigenvalue $-e^{-|\vec{k}|d}/2$. As $\DD$ is diagonal in
this orthonormal basis, then its singular values are the magnitudes of
the eigenvalues, so $g(\vec{k}) = e^{-|\vec{k}|d}/2$. Slight care must
be taken with respect to the orthogonality term $(2\pi)^{2} \delta^{2}
(\vec{k} - \vec{k}')$, as $(2\pi)^{2} \delta^{2}_{\vec{k}} (0) =
\int_{-\infty}^{\infty} \int_{-\infty}^{\infty} \mathrm{d}x~\mathrm{d}y =
A$. Knowing this, it can be seen that $\sum_{i} \to A \iint
\frac{\mathrm{d}^{2} k}{(2\pi)^{2}}$, so plugging $g(\vec{k})$ into
the various bounds gives the analytical expressions in the main text.

The derivation of the singular values of $\Gvac_{\mathrm{BA}}$ for
extended slabs of finite thickness is similar to that for
semi-infinite thickness. In particular (dropping the
$\mathbb{M}^{\mathrm{s}}$ term and evaluating all terms in the
nonretarded approximation), the operator
\begin{multline}
  \DD (\vec{k}, \vec{k}', z, z') = -\frac{\im\omega^{2}}{2c^{2}}
  \mathbb{M}^{\mathrm{p}} e^{-|\vec{k}|(z + z')} \times (2\pi)^{2}
  \delta^{2} (\vec{k} - \vec{k}') \times \\ \Theta(z - d/2) \Theta(z'
  - d/2) \Theta(h + d/2 - z) \Theta(h + d/2 - z')
\end{multline}
has a tensor term $-\frac{\im\omega^{2}}{2c^{2}}
\mathbb{M}^{\mathrm{p}}$ which can be written as the Cartesian outer
product $-|\vec{k}|((-\im\vec{k} +
|\vec{k}|\vec{e}_{z})/(\sqrt{2}|\vec{k}|)) \otimes ((\im\vec{k} +
|\vec{k}|\vec{e}_{z})/(\sqrt{2}|\vec{k}|))$. The spatial term
$e^{-|\vec{k}|(z + z')}$ under the new spatial domain of finite
thickness $h$ satisfies $\int_{d/2}^{d/2 + h} e^{-|\vec{k}|(z + z'')}
e^{-|\vec{k}|(z'' + z')}~\mathrm{d}z'' = ((e^{-|\vec{k}|d} -
e^{-|\vec{k}|(d + 2h)})/(2|\vec{k}|))e^{-|\vec{k}|(z +
  z')}$. Therefore, this operator may be written as the outer product,
\begin{multline}
  \DD(\vec{k}, \vec{k}', z, z') = -\frac{e^{-|\vec{k}|d}(1 -
    e^{-2|\vec{k}|h})}{2} \times \\ \vec{v}^{(0)} (\vec{k}, z) \otimes
  \vec{v}^{(0)\star} (\vec{k}, z') \Theta(z - d/2) \Theta(z' - d/2)
  \\ \Theta(h + d/2 - z) \Theta(h + d/2 - z') (2\pi)^{2} \delta^{2}
  (\vec{k} - \vec{k}')
\end{multline}
having defined the new plane-wave eigenfunctions,
\begin{equation}
  \vec{v}^{(0)} (\vec{k}, z) = (|\vec{k}|(1 -
  e^{-2|\vec{k}|h}))^{-1/2} (-\im\vec{k} +
  |\vec{k}|\vec{e}_{z})e^{-|\vec{k}|(z - d/2)}
\end{equation}
normalized such that $\int_{d/2}^{h + d/2} \vec{v}^{(0)\star}
(\vec{k}, z) \cdot \vec{v}^{(0)} (\vec{k}, z)~\mathrm{d}z = 1$, with
corresponding eigenvalue $-e^{-|\vec{k}|d} (1 -
e^{-2|\vec{k}|h})/2$. The corresponding singular values are therefore,
\[g(\vec{k}) = \frac{e^{-|\vec{k}|d}}{2}(1 - e^{-2|\vec{k}|h}).\]
We note that when evaluating $\Phi_{\mathrm{opt}}$, the transition
between the contributions that do or do not saturate the Landauer
bound corresponds to the condition $\sqrt{\zeta_{\mathrm{A}}
  \zeta_{\mathrm{B}}} e^{-|\vec{k}|d} (1 - e^{-2|\vec{k}|h})/2 = 1$,
so the corresponding value of $|\vec{k}|$ must be determined by
numerically solving this transcendental equation; such a solution will
only exist for a given $\eta = h/d$ if $\sqrt{\zeta_{\mathrm{A}}
  \zeta_{\mathrm{B}}} > \eta^{-1} (1 + 2\eta)^{1 + 1/(2\eta)}$, and if
this condition is violated, then the integrand $\frac{2}{\pi}
\frac{\zeta_{\mathrm{A}} \zeta_{\mathrm{B}}
  (g(\vec{k}))^{2}}{(\zeta_{\mathrm{A}} \zeta_{\mathrm{B}}
  (g(\vec{k}))^{2})^{2}}$ must be used for all $\vec{k}$.

\nocite{apsrev41Control} \bibliographystyle{apsrev4-1}

\bibliography{nearfieldheatboundspaper}
\end{document}